\documentclass{moriond}

\def\Journal#1#2#3#4{{#1} {\bf #2}, #3 (#4)}

\def\PLB{{\em Phys. Lett.}  B}

\def\be{\begin{equation}}
\def\ee{\end{equation}}
\def\bea{\begin{eqnarray}}
\def\eea{\end{eqnarray}}

\usepackage{float}
\usepackage{graphicx} 
\usepackage{amsmath}

\newcommand{\Photo}{}

\newcommand{\PS}{{\rm PS}}

\begin{document}
\vspace*{4cm}
\title{Flavour non-universal Pati-Salam unification and neutrino masses}

\author{Julie~Pag\`es}

\address{Physik-Institut, Universit\"at Z\"urich, CH-8057 Z\"urich, Switzerland}

\maketitle
\abstracts{
The B anomalies, by their distinctive flavour structure \textit{i.e.} $U(2)^5$, bring a new piece to the long-standing flavour puzzle. The three-site flavour non-universal Pati-Salam (PS) model, which unifies quarks and leptons, provides through the ratios of vacuum expectation values (VEV) acquired at different scales, a combined explanation of the charged and neutral current B anomalies as well as of the mass hierarchies of the Standard Model (SM). The mixings, in new, flavour non-universal, gauge interactions, as well as in the Yukawa couplings, are obtained through suppressed nearest-neighbour interactions.
In this three-site model context, the inverse seesaw mechanism is realised, with the minimal addition of three fermion singlets $S_L^{(i)}$ and where $U(1)_F$ fermion number is broken dynamically via new singlet scalars $\Phi_i$,  yielding an anarchic light neutrino mass matrix in consistency with data, despite the $U(2)^5$ flavour symmetry observed in the Yukawa matrices. A prediction of this model is Pontecorvo-Maki-Nakagawa-Sakata (PMNS) unitarity violation with a $33$ entry close to experimental bounds.
The full model finds a natural 5D interpretation with three (almost equidistant) defects in a warped extra dimension, where the exponential hierarchies in VEV ratios of the 4D Lagrangian arise from O(1) differences in the 5D field bulk masses. \\This proceeding is based on arXiv:2012.10492.}

\section{Introduction}
On the one hand, we have the B anomalies. They are a consistent and persistent set of discrepancies between the SM predictions and the experimental results in semileptonic and rare decays of B mesons. They are classified into two categories: \textit{i}) deviations from $\tau/\mu,e$ universality in $b\to c\ell\bar\nu$ charged-current transitions, and \textit{ii}) deviations from $\mu/e$ universality in $b\to s\ell^+\ell^-$ neutral-current transitions. The global significance of the new physics hypothesis in the $b\to s\ell^+ \ell^-$ system is 3.9$\sigma$ \cite{Lancierini:2021} and it exceeds 3.1$\sigma$ in the $b\to c\ell \bar \nu$ system. 
An effective fields theory (EFT) analysis of the contribution required to explain the anomalies revealed that the new mediator should be relatively light ($\sim$ TeV) and coupled mainly to the third family of quarks and leptons. The perfect candidate for a combined (charged- and neutral-currents) explanation was identified to be the $U_1 \sim (3,1)_{2/3}$ vector leptoquark (LQ) \cite{Buttazzo:2017}, with non-universal flavour couplings.

On the other hand, we have the flavour puzzle. It is a structural problem of the SM: why is the SM flavour sector so complex with 13 (+9 for neutrinos) masses and mixings spanning 5 orders of magnitude, compared to the gauge sector which can describe all interactions with only 3 (similar order) couplings?

In order to construct a theory of flavour, also capable to account for the deviations observed in the B mesons decays, identifying common features helps selecting model building directions. This is why the use of flavour symmetries is especially relevant in this case. It has already been noted that the SM Yukawa couplings respect an approximate $U(2)^5 = U(2)_q \times U(2)_\ell \times U(2)_u \times U(2)_d \times U(2)_e$ flavour symmetry \cite{Barbieri:2011} where only the third family quarks and leptons have a mass, while the light families mixings and masses can be parametrised by small $U(2)^5$ spurions. Interestingly the same flavour symmetry yields the right flavour structure of the LQ interactions in order to address the B anomalies \cite{Buttazzo:2017,Fuentes-Martin:2019,Barbieri:2015}.

In these proceedings, we follow ref. \cite{Fuentes-Martin:2020} by first reviewing in section \ref{PS3model} the Pati-Salam cubed model \cite{Bordone:2017} and discuss its interpretation in terms of a deconstructed 5 dimensional model. At low energy, we witness the emergence of an accidental $U(2)^5$ flavour symmetry from the construction, which gives its structure to the Yukawa and the LQ couplings. In section \ref{Neutrinoext} the model is extended to account for the smallness and non-hierarchical nature of the neutrino masses. Finally in section \ref{5Dmodel} the potential UV completion from a continuous 5D model is presented.

\section{PS$^3$ model} \label{PS3model}

The gauge group of Pati-Salam unification \cite{Pati:1974} is the minimal group containing the SM and the $U_1$ leptoquark. 
In order to introduce flavour non-universality in its coupling to matter, the model is made of three sites
\begin{equation}
	{\PS}^3 \equiv \PS_1 \times \PS_2 \times \PS_3\,,
\end{equation}
with $ \PS_i = SU(4)_i \times SU(2)_{L,i} \times SU(2)_{R_i}$ and where
each fermion family is only charged under one of the $\PS_i$. They are described by the fields
\begin{equation}
	\Psi_L^{(i)}=\begin{pmatrix}
		u_L & d_L\\
		\nu_L & \ell_L
	\end{pmatrix}^{(i)}\sim(4,2,1)_i\,, \qquad \Psi_R^{(i)}=\begin{pmatrix}
		u_R & d_R\\
		\nu_R & \ell_R
	\end{pmatrix}^{(i)} \sim (4,1,2)_i\,. 
\end{equation} 
The three-site $\PS^3$ can be viewed as the 4D deconstruction of a 5D $\PS$ model with 3 special points (defects) along the 5$^{\rm th}$ dimension. In this context, the fermion families could be originating from the same 5D fields but different zero-modes, each mostly localised on one of the special points, with a tail overlapping with the other sites parametrised by $\epsilon_{ij}^L$  (see fig. \ref{fig:5Dpicture} and eq. \eqref{eq:Diracmass}).

To understand how this UV model affects the low energy observables, its extended gauge sector must undergo several spontaneous symmetry breakings (SSB). The different SSB fall into two categories:
\begin{itemize}
	\item Vertical breaking (happening independently on each site)
	\begin{equation}
		\PS_i~ \xrightarrow{\langle \Sigma_i\rangle}~ {\rm SM}_i ~\xrightarrow{\langle H_i \rangle} ~SU(3)_i \times U(1)_i
	\end{equation}
	where the scalars responsible are a new field $\Sigma_i\sim(4,1,2)_i$ and an extended Higgs field $H_i\sim(1,2,\bar 2)_i$ which contains two SM Higgses.
	\item Horizontal breaking (connecting the different sites)
	\begin{equation}
		G_i \times G_j \xrightarrow{\langle \Omega_{ij} \rangle} G_{\rm diagonal}
	\end{equation}
	with three non-linear link fields $\Omega^4_{ij} \sim (4,1,1)_i \times (\bar 4,1,1)_j$, $\Omega^L_{ij} \sim (1,2,1)_i \times (1,\bar 2,1)_j$ and $\Omega^R_{ij} \sim (1,1,2)_i \times (1,1,\bar 2)_j$, one for each group of PS.
\end{itemize}
The non-linear nature of the horizontal breaking scalars suggest a UV cut-off of the 4D theory at $\Lambda \sim 4 \pi \langle \Omega_{ij}^{4,L,R} \rangle$, where another UV picture is required, \textit{e.g.} the continuous 5D model.
Indeed, while the scalars responsible for the vertical breaking have to be extra fields of the model both in the 4D three-site model and in the 5D model, the non-linear link fields can be seen as the 5$^{\rm th}$ component of the Kaluza-Klein (KK) modes of the 5D gauge bosons.

\begin{figure}[H]
\centering
\includegraphics[width=10.5cm]{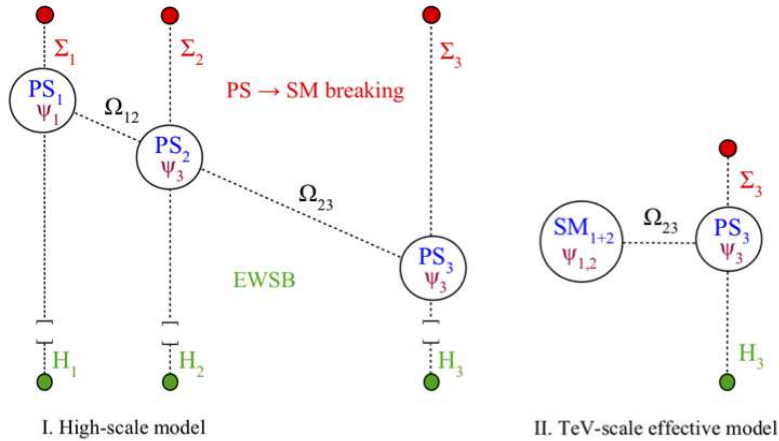}   
\caption{Schematic representation of the SSB structure in $\PS^3$. Long(short) lines indicate small(large) non-vanishing VEV.}
\label{3site}
\end{figure}

In the discrete picture, the Yukawa hierarchies and mixings are generated by terms of the kind
\begin{equation} \label{eq:Diracmass}
	-\mathcal{L}_Y \supset y_i \bar \Psi_l^{(i)} H_i \Psi_R^{(i)} + y_{23} \epsilon_{23}^L \bar \Psi_L^{(2)} \frac{\Omega_{23}^4}{\langle\Omega_{23}^4 \rangle} \frac{\Omega_{23}^L}{\langle\Omega_{23}^L \rangle} H_3 \Psi_R^{(3)} + ...
\end{equation}
with $O(1)$ coefficients $y_i$ and $y_{ij}$.
Stringent flavour constraints requires the VEV of $\Sigma$ to be mostly localised on the first site, while the hierarchies observed in the SM Yukawa couplings dominantly localise the Higgs VEV on the third site.
The suppressed nearest-neighbour interactions, which also provides stability to the SM Higgs sector as illustrated with a toy model in \cite{Allwicher:2020},
 can be described by the ratio of VEVs on the different sites,
\begin{equation}
	 \epsilon^\Sigma_{23} \epsilon^\Sigma_{12} \langle \Sigma_1 \rangle =  \epsilon^\Sigma_{23} \langle \Sigma_2  \rangle = \langle \Sigma_3 \rangle\,, \qquad 
	 \langle H_1 \rangle =\epsilon^H_{12} \langle H_2 \rangle = \epsilon^H_{12} \epsilon^H_{23} \langle H_3 \rangle
\end{equation} 
which mimic the spread of the VEV profile in the 5D model (see fig. \ref{5Dmodel}).
Controlling these VEVs, as illustrated in fig. \ref{3site}.I (taken from ref. \cite{Fuentes-Martin:2020}), allows us to reproduce the flavour structure of the SM. 

The limits on flavour-changing neutral currents involving the first two generations, \textit{e.g.} $K$-$\bar K$ and $D$-$\bar D$ mixing, $K_L\to \mu e$, set a lower bound at $10^3$ TeV for the VEVs of $\Sigma_{1,2}$ and $\Omega^{4,L,R}_{12}$.
Below this scale, the effective model, presented in fig. \ref{3site}.II, exhibits an accidental $U(2)^5$ flavour symmetry, whose spurions can be identified with the nearest-neighbour suppression factors $\epsilon_{ij}^H$ and  $\epsilon_{ij}^L$. This approximate flavour symmetry provides protection for the light families against the new gauge sector, in $\PS_3$, so that the VEVs of $\Sigma_{3}$ and $\Omega^{4,L,R}_{23}$ can be as low as a few TeV.
The leptoquark obtained from the breaking of $SU(4)_3$ in $\PS_3$ ticks all the boxes to explain the B anomalies \cite{DiLuzio:2018,Cornella:2019,Fuentes-Martin:2020loop}: TeV-scale, coupled mainly to 3$^{\rm rd}$ family, with suppressed interaction to lighter generations ($\propto \epsilon_{ij}^L$).

The three-site $\PS$ model with nearest-neighbour interactions provides a dynamical description of the SM Yukawa couplings consistent with an explanation for the B anomalies, assuming the approximate relations
\begin{equation} \label{eq:epsQuarks}
	\epsilon_{ij}^H \sim (\epsilon_{ij}^L)^2 \sim 10^{-2}\,, \qquad \epsilon_{12}^L \approx 2 \epsilon_{23}^L\,.
\end{equation}

\section{Neutrino extension} \label{Neutrinoext}

In its original formulation, the $\PS^3$ model entails a pathological neutrino sector because of two reasons:
\begin{itemize}
	\item quark-lepton unification implies Dirac masses with $m_e^{(i)} =m _d^{(i)}$ and $m_\nu^{(i)}= m_u^{(i)}$;
	\item low-scale unification, $\langle \Sigma_3 \rangle \sim$ TeV,  prevents us from writing a Majorana mass for the right-handed neutrino $m_{\nu_R}^{(3)}$ above this scale (for type-I seesaw $m_\nu \approx m_D^2/m_{\nu_R}$ where $m_D$ is the Dirac mass of the neutrino).
\end{itemize}

Most neutrino mechanisms thus fail to explain the smallness of their mass. A new source is needed to suppress a Dirac mass $m_D^{(3)}\approx m_t \sim 100$ GeV which is 12 orders of magnitude above the experimental bound from cosmology $m_\nu^{\rm exp} \leq 0.1$ eV.

The minimal extension consists in adding three fermion singlets $S_L^{(i)}$ with Majorana masses to implement the inverse seesaw mechanism \cite{Greljo:2018}.
The inverse seesaw mechanism is made of 3 scales: 
a Dirac mass $m_D$ between $\nu_L$ and $\nu_R$ given by eq. \eqref{eq:Diracmass};
a Dirac mass $m_R$ between $\nu_R$ and $S_L$ generated when $\Sigma_i$ acquire a VEV from terms of the type
\begin{equation}
	-\mathcal{L}_\nu \supset y_i^\Sigma \bar S_L^{(i)} \Sigma_i^\dagger \Psi_R^{(i)} + y_{21}^\Sigma \epsilon_{21}^S \bar S_L^{(2)} \Sigma_1^\dagger \Psi_R^{(1)} + ... \,,
\end{equation}
with $O(1)$ coefficients $y_i^{\Sigma}$ and ${y_{ij}^\Sigma}$, and with $\epsilon_{ij}^S$ analogous to $\epsilon_{ij}^L$ but for the fermion singlets $S_L^{(i)}$ instead of $\Psi_L^{(i)}$;
and a Majorana mass $\mu$ for $S_L$.
The diagonalization of the $9\times9$ mass matrix written in the Lagrangian as $\bar n M_\nu n$ in the basis $n=\begin{pmatrix}
	\nu_L^c & \nu_R & S_L^c \end{pmatrix}^T$
\begin{equation}
	M_\nu = \begin{pmatrix}
	0 & m_D & 0 \\
	m_D^T & 0 & m_R^T \\
	0 & m_R & \mu 
	\end{pmatrix}
\end{equation}
yields at leading order in $m_D/m_R$ a light neutrino Majorana mass matrix given by
\begin{equation} \label{eq:Neutrinomass}
	m_\nu \approx m_D m_R^{-1} \mu (m_R^{-1})^T m_D^T\,.
\end{equation}
Note that in this expression the $U(2)^5$ flavour symmetry, required for the quark sector, feeds back quadratically with the ratio $m_D^{(i)}/m_R^{(i)}$. A very hierarchical structure of the Majorana mass $\mu$ is thus needed to obtain an anarchical neutrino mass matrix $m_\nu$. 
This is achieved dynamically by the addition of a singlet scalar $\Phi_i$ breaking spontaneously $U(1)_F$ fermion number \footnote{Since $U(1)_F$ breaking happens only in the leptonic sector, baryon number remains as an accidental global symmetry. The Goldstone from this SSB called the Majoron can be given a mass via explicit symmetry breaking \cite{Fuentes-Martin:2020loop}.}
 with a VEV mostly localised on the first site and suppressed on the other sites

\begin{equation}
	-\mathcal{L}_\nu \supset \frac{y_i^\Phi}{2} \bar S_L^{\scriptsize (i)} \Phi_i {S_L^{(i)}}^c \qquad \text{with} \qquad 	 \epsilon^\Phi_{23} \epsilon^\Phi_{12} \langle \Phi_1 \rangle =  \epsilon^\Phi_{23} \langle \Phi_2  \rangle = \langle \Phi_3 \rangle\,,
\end{equation}
similarly to $\Sigma$ but with a steeper suppression/profile (see fig. \ref{fig:5Dpicture}).

In order to obtain a fully anarchical mass matrix $m_\nu$, also in the off-diagonal entries, allowing for arbitrarily large angles in the PMNS, the suppressed nearest-neighbour interaction parameters must satisfy
\begin{equation} \label{eq:epsLeptons}
	\epsilon_{ij}^S \sim \epsilon_{ij}^H \epsilon_{ij}^\Sigma \qquad \text{and} \qquad \epsilon_{ij}^\Phi \sim (\epsilon_{ij}^S)^2\,.
\end{equation}

Interestingly, this three times inverse seesaw model does not appear the same on each site. Indeed, the first site relation can be simplified assuming a unique VEV for the fields localised on the first site \textit{i.e.} $\langle \Sigma_1 \rangle =\langle \Phi_1 \rangle \sim \Lambda_{\rm UV}$, 
\begin{equation}
	m_\nu^{(1)}= \frac{m_u^2}{m_t^2} \frac{v^2}{\langle \Sigma_1 \rangle^2}\langle \Phi_1 \rangle \to \frac{m_u^2}{m_t^2} \frac{v^2}{\Lambda_{\rm UV}}\,,
\end{equation}
meaning that for the first site, the squared Higgs suppression $\sim 10^{-8}$ and the $U(1)_F$ breaking scale $\sim 10^4$ TeV are high enough so that a standard type-I seesaw can realise the small neutrino mass.

A clear signature of this neutrino extension is due to the mixing between the active neutrino and the pseudo-Dirac heavy neutral leptons, giving rise to PMNS unitarity violation with a specific pattern given by 
\begin{equation}
	\eta \equiv |1-NN^\dagger| \sim \left| \frac{m_D^{(3)}}{m_R^{(3)}} \right|^2 \begin{pmatrix}
		(\epsilon_{12}^L\epsilon_{23}^L)^2 & \epsilon_{12}^L(\epsilon_{23}^L)^2 & \epsilon_{12}^L\epsilon_{23}^L \\
		\epsilon_{12}^L(\epsilon_{23}^L)^2 & (\epsilon_{23}^L)^2 & \epsilon_{23}^L \\
		\epsilon_{12}^L\epsilon_{23}^L & \epsilon_{23}^L & 1  \\
	\end{pmatrix}
\end{equation}
where $N$ is the non-unitary PMNS matrix. As can be seen, the largest violation of unitarity happens in the bottom right entry that we estimate at
\begin{equation}
	\eta_{33} \approx  \left| \frac{m_D^{(3)}}{m_R^{(3)}} \right|^2 \sim \left| \frac{100 \rm ~GeV}{2 \rm ~TeV} \right|^2 = 2.5 \times 10^{-3}\,,
\end{equation}
very close to the experimental bound $\eta_{33}^{\rm exp} < 5.3 \times 10^{-3} [90\% \rm ~C.L.]$ found in ref. \cite{Antusch:2014}. It is the most stringent bound since the other entries are protected by $\epsilon_{ij}^L$ suppressions.

\section{5D model} \label{5Dmodel}

\begin{figure}[H]
\centering
\includegraphics[width=10.4cm]{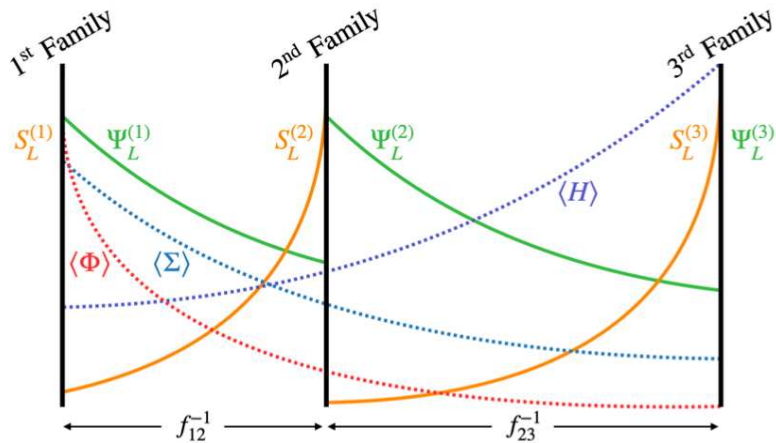}   
\caption{Scalars VEVs and fermions zero-mode profiles along the $5^{\rm th}$ dimension in the continuous model, $\PS_{\rm 5D}$. Right-handed fields (not shown) are assumed as fully localised on their respective site.}
\label{fig:5Dpicture}
\end{figure}

The three-site model with suppression factors $\epsilon_{ij}^H,\epsilon_{ij}^\Sigma,\epsilon_{ij}^\Phi,\epsilon_{ij}^L,\epsilon_{ij}^S$ from nearest-neighbour interactions finds a natural interpretation in terms of a 5D model for flavour with 3 special points. 
Each fermion zero-mode is mostly localised on one of the site with an exponentially decaying profile extending to the other sites.
Each scalar is broken first on only one of the sites with an exponentially suppressed propagation of the VEV to the other sites.
The full picture is illustrated in fig. \ref{fig:5Dpicture}.
In both cases, since 4D couplings are given by the overlap of the profiles and right-handed fermion field are assumed fully localised on each site, we can identify the exponential profile with the nearest-neighbour suppression factors from the three-site model
\begin{equation}
	\epsilon^F_{ij} = e^{-M_F/f_{ij}} 
\end{equation}
where $f_{ij}$ is the inverse of the distance between site $i$ and $j$, and $M_F$ is the 5D bulk mass of the corresponding field $F$.
This identification fixes the matching between the UV complete 5D picture and the 4D deconstructed picture with the flavour and neutrino  relations, respectively eq. \eqref{eq:epsQuarks} and \eqref{eq:epsLeptons},
translated into
\begin{equation}
	M_H \approx 2 M_L\,, \quad M_\Phi \approx 2 M_S\,, \qquad M_\Sigma \approx M_S -2 M_L
\end{equation}
The advantage of these 5D relations is that same order 5D bulk masses $M_F$ can explain the big ratios of VEVs in the discrete model and nearest-neighbour exponential suppressions. 

The flavour and neutrino sector finds a nice interpretation in this 5D construction. 
Regarding the gauge sector, dangerous KK modes coupling to the light families can be avoided by considering a warped extra dimension. This  direction is being explored in the last part of \cite{Fuentes-Martin:2020} and in future works.

\section*{Acknowledgments}
I would like to thank the Organisers of Les Rencontres de Physique de la Vall\'ee d'Aoste for the invitation, as well as my colleagues Javier Fuentes-Mart\`in, Gino Isidori and Ben Stefanek for the collaboration on this work.

\end{document}